\documentstyle [12pt]{article}
\begin{document}

\rightline{FTUV 93-36}
\rightline{August 1993}
\vspace{3\baselineskip}

\begin{center}
{\LARGE{\bf Reflection equations and q-Minkowski space algebras}}
\vspace{0,3cm}
\end{center}
\vspace{4\baselineskip}

\begin{center}
{\large{J.A. de Azc\'{a}rraga, P.P. Kulish\footnote{On
leave of absence from the St.Petersburg's Branch of the Steklov Mathematical
Institute of the Russian Academy of Sciences.} and F. R\'{o}denas}}
\end{center}
\vspace{0,5cm}
\addtocounter{footnote}{-1}

\begin{center}
{\small{\it Departamento de F\'{\i}sica Te\'{o}rica and IFIC,\\
Centro Mixto Universidad de Valencia-CSIC\\
46100-Burjassot (Valencia), Spain.}}
\end{center}
\vspace{1cm}

\begin{abstract}
We express the defining relations of the $q$-deformed Minkowski space
algebra as well as that of the corresponding derivatives
and differentials in the form of
reflection equations. This formulation encompasses the covariance
properties with respect the quantum Lorentz group action in a straightforward
way.
\end{abstract}

The $q$-deformation of the Lorentz group and
Minkowski space has been the subject of active
research in the last few years \cite{8,9,10,11,12}.
Its interest stems from the fact that  any
physical application of quantum groups to problems related with
spacetime symmetries requires an understanding of the possible
non-commutative geometry of spacetime.
This demands that the action (coaction) of the quantum Lorentz
group ${\cal L}_q$  on the $q$-Minkowski space
${\cal M}_q$  is defined in a way which
is consistent with its non-commuting properties. This letter is
devoted to showing that the reflection equations (RE)
(see \cite{1,2,3,4,5} and refs. therein) provide the adequate
general framework for such a purpose, from which the commutation
relations among coordinates and derivatives (and some other properties)
can be extracted easily. This is because reflection equations, being
a consistent extension of the Yang-Baxter equation (YBE)

\begin{equation} \label{one}
R_{12} (u) R_{13} (u+v) R_{23}(v)=R_{23}(v) R_{13}(u+v) R_{12}(u) \; ,
\end{equation}

\noindent
give rise to algebraic structures closely related with quantum groups.

In this paper, as in the theory of quantum groups corresponding to simple Lie
groups and algebras, we will consider only constant solutions to the
Yang-Baxter equation (\ref{one}) which do not
depend on the spectral parameter $u$. Hence the
reflection equations will also be spectral parameter independent
\cite{2,3,4,5}

\begin{equation} \label{two}
R^{(1)} K_{1} R^{(2)} K_{2} = K_{2} R^{(3)} K_{1} R^{(4)},
\end{equation}

\noindent
where $R^{(i)}$ are appropriate constant solutions of the YBE (\ref{one}).
Having in mind the application of reflection
equations to $q$-deformed Minkowski spaces we
restrict ourselves to the case where all four $R$-matrices in (\ref{two}) are
expressed in terms of the quantum group $GL_{q}(2)$ $R$-matrix \cite{6}

\begin{equation} \label{three}
\left(
\begin{array}{cccc}
q & 0 & 0 & 0 \\
0 & 1 & 0 & 0 \\
0 & \lambda & 1 & 0 \\
0 & 0 & 0 & q \\
\end{array}
\right)
\,,  \quad \quad \lambda = q - q^{-1} \quad .
\end{equation}

The quantum group theory gives rise to an increasing number of explicit
examples
of non-commutative geometry (see e.g. \cite{7}). In particular the important
physical question of the  $q$-deformation of
the Lorentz group and Minkowski space is discussed in a few
papers (see \cite{8,9,10,11,12}
and references therein). The starting point of the latter
ones is the relation of the Lorentz group ${\cal L}$ to $SL(2,C)$
\cite{8} and the spinorial construction
of the Minkowski space coordinates $x^{\mu}$ \cite{9,10,11,12}.
We will show that it is possible to use for the definition of a
$q$-deformed Minkowski space the well-known matrix relation expressing the
$2 \rightarrow 1$ homomorphism between $SL(2,C)$ and the Lorentz group,

\begin{equation} \label{four}
x^{\mu} \mapsto \Lambda^{\mu}_{\nu} x^{\nu} \;, \qquad \sigma_{\mu} x^{\mu}
\mapsto \sigma_{\mu} x'^{\mu}= A \sigma_{\mu} x^{\mu} A^{\dagger}\;,
\quad A,A^\dagger \in SL(2,C),
\end{equation}

\noindent
in the framework of the $R$-matrix formalism \cite{6,3} and reflection
equations, translating in this way to the deformed case the covariance
properties of $\sigma_{\mu} x^{\mu}$ expressed by (\ref{four}).
In this manner, the statement of \cite{4} on the equivalence of the
$sl_{q}(2)$-reflection equation algebra and the $q$-Minkowski space algebra
${\cal M}_{q}$
\cite{9} is demonstrated explicitly. By extending the method to the
algebra of derivatives and differentials we shall be able to find
compact, simple expressions for their respective commutation relations in a
natural way, clarifying the source of the possible ambiguities in them
as well as the relations among the results of different authors.

In order to $q$-deform the transformation (\ref{four}) we propose
to consider instead
of $\sigma_{\mu} x^{\mu}$ a 2$\times$2  matrix $K$, the entries of which are
the generators of the $q$-Minkowski space algebra in question. Following
\cite{9,11} we
introduce two isomorphic but mutually non-commuting copies of
the quantum group $SL_{q} (2,C)$. The commutation relations among these
quantum
group generators ($a, b, c, d$; $\tilde{a}, \tilde{b}, \tilde{c}, \tilde{d}$)
in matrix form \cite{9,12} look like this:

\begin{equation} \label{five}
R_{12} M_{1} M_{2} = M_{2} M_{1} R_{12}
\end{equation}
\begin{equation}\label{six}
R_{12} \tilde{M}_{1} \tilde{M}_{2} = \tilde{M}_{2} \tilde{M}_{1} {R}_{12}
\end{equation}
\begin{equation}\label{seven}
R_{12} M_{1} \tilde{M}_{2} = \tilde{M}_{2} M_{1} R_{12}
\end{equation}

\noindent
where

$$
M=
\left(
\begin{array}{cc}
a & b \\
c & d \\
\end{array} \right)
 , \qquad   a d - q b c = 1 \; , \quad q \in {\bf R}\,,
$$

\noindent
and $\tilde{M}$ is used for an isomorphic copy of $SL_{q} (2,C)$
with generators
$\tilde{a}, \tilde{b}, \tilde{c}, \tilde{d}\;$\, $(\tilde{a} \tilde{d} -
q \tilde{b}
\tilde{c} = 1)$. We use the standard notations
for the YBE and the quantum group theory (cf. \cite{6,7}) {\it e.g.}
$M_1$=$M \otimes I$, $M_2$=$I \otimes M$ are 4$\times$4 matrices in
${\bf C}^2$$\otimes$${\bf C}^2$.

The transformation of the generators $(\alpha, \beta, \gamma, \delta)$ of a
$q$-Minkowski space algebra ${\cal M}_{q}$
is written as (cf. eq. (\ref{four}))

\begin{equation}\label{eight}
\phi: K=
\left(
\begin{array}{cc}
\alpha & \beta \\
\gamma & \delta \\
\end{array}
\right)
\longmapsto K' = M K \tilde{M}^{-1}
\end{equation}

\noindent
where it is assumed
that the entries of $K$ commute with those of $M$ and $\tilde{M}$.
If the reality condition $\tilde{M}^{-1}= M^{\dagger}$
(which in the $q$=1 case gives $(D^{0,\frac{1}{2}})^{-1} =(D^{\frac{1}{2},0})^
{\dagger}$ for the two fundamental representations of the Lorentz group)
is imposed on (\ref{eight}), then this equation realizes
the action\footnote{We use freely the word `action', although in the
Hopf algebra theory the map $\phi: {\cal M}_q \rightarrow {\cal L}_q \otimes
{\cal M}_q$ is referred to as a coaction.}
of the $q$-Lorentz group ${\cal L}_q$
on the entries of $K$ much in the same way as for the
$q$=1 case this equation implements the classical homomorphism
$SL(2,C)$$\rightarrow {\cal L}$. The action (coaction) (\ref{eight})
is extended to any element of the Minkowski algebra ${\cal M}_{q}$
by the requirement of being an algebra homomorphism ({\it e.g.}
$\phi (\alpha \beta)= \phi (\alpha) \phi (\beta)$).
The next important
requirement is the covariance property of the algebra ${\cal M}_{q}$ with
respect to (\ref{eight}). This means that
transformed generators $\alpha',...,\delta'$, {\it e.g.}

\begin{equation}\label{nine}
\phi (\alpha)=
\alpha' = a \tilde{d} \alpha + b \tilde{d} \gamma - q a \tilde{c} \beta - q b
\tilde{c} \delta \; ,
\end{equation}

\noindent
{\it etc},
must satisfy the same relations as the original ones. Such relations among
$\alpha,...,\delta$ are given by the reflection equation (\ref{two})
with appropriate matrices $R^{(i)}$ \cite{2,4}:

\begin{equation}\label{ten}
R_{12} K_{1} R_{21} K_{2} = K_{2} R_{12} K_{1} R_{21},
\end{equation}

\noindent
where

\begin{equation}\label{eleven}
R_{21} = {\cal P} R_{12} {\cal P}
\end{equation}

\noindent
and ${\cal P}$ is the permutation operator. In
${\bf C}^{2}$$\otimes$${\bf C}^{2}$, in the  basis
$(e_{1} \otimes e_{1}, e_{1} \otimes e_{2}, e_{2} \otimes e_{1}, e_{2} \otimes
e_{2})$ it is given by

$$
\left(
\begin{array}{cccc}
1 & 0 & 0 & 0 \\
0 & 0 & 1 & 0 \\
0 & 1 & 0 & 0 \\
0 & 0 & 0 & 1 \\
\end{array}
\right)
\;.
$$

The identification of the Minkowski
algebra ${\cal M}_q$ with the associative
reflection equation algebra (REA) defined by the relations (\ref{ten})
guarantees the covariance of the commuting properties of the
$q$-Minkowski `coordinates' with respect to the coaction (\ref{eight})
(i.e., with respect to ${\cal L}_q$).
In terms of the entries  of $K$ \cite{2,4,13}
($\lambda$=$q-q^{-1}$) the commutation relations read

\begin{equation}\label{twelve}
\begin{array}{rcl}
\alpha \gamma  & = & q^{2} \gamma \alpha \; ,\\
    & &  \\
\alpha \beta &  = &  q^{-2} \beta \alpha \;, \\
  &  &   \\
\alpha \delta & =& \delta \alpha \;, \\
\end{array} \qquad
\begin{array}{ccl}
{} [\gamma, \delta] & = &  (\lambda/q) \gamma \alpha  \; , \\
    &   &   \\
{} [\beta, \gamma] &=& (\lambda/q) (\alpha \delta - \alpha^{2}) \; ,  \\
    &  &  \\
{} [\beta, \delta] &=& \; -  (\lambda/q) \alpha \beta \;.
\end{array}
\end{equation}

\noindent
To see explicitly that equation (\ref{ten}) is invariant under the
coaction (\ref{eight}) it is sufficient to write (\ref{ten}) for
$K'$=$MK\tilde{M}^{-1}$ and then check, using the defining relations
(\ref{five}-\ref{seven}), that the additional factors $M_1$,
$M_2$, $\tilde{M}^{-1}_1$, $\tilde{M}^{-1}_2$ cancel out
reproducing again eq. (\ref{ten}) for $K$.

The centrality of the following two elements of ${\cal M}_{q}$

\begin{equation}\label{thirteen}
c_{1} \equiv \alpha + q^{2} \delta \; ,
\end{equation}
\begin{equation}\label{fourteen}
c_{2} \equiv \alpha \delta - q^{2} \gamma \beta \;,
\end{equation}

\noindent
easily follows from the invariance property of the $q$-trace \cite{6,7,5}.

\begin{equation}\label{fifteen}
tr_{q} K \equiv  tr D K = \alpha + q^{2} \delta \;, \quad
\qquad D = diag (1, q^{2})\; ,
\end{equation}

\noindent
with respect to the quantum group coaction

\begin{equation}\label{sixteen}
tr_{q} K = tr_{q} \{ M K M^{-1} \} \; .
\end{equation}

\noindent
This is easily  seen from the transformed RE (\ref{ten}),

\begin{equation}\label{seventeen}
K_{2} R_{12} K_{1} R^{-1}_{12} = R^{-1}_{21} K_{1} R_{21} K_{2}\;,
\end{equation}

\noindent
taking into account that $R_{12}$ and  $R_{21}^{-1}$ give the representations
of the quantum group (\ref{five}) as 2$\times$2 matrices in the first space of
${\bf C}^{2}$$\otimes$${\bf C}^{2}$ and
taking the $q$-trace of (\ref{seventeen}) with respect to
the first space:

\begin{equation}\label{eighteen}
K c_{1}= c_{1} K
\end{equation}

\noindent
(commutativity of $c_{1}$ with all generators of ${\cal M}_{q}$).
As for $c_2$, it can also be expressed in terms
of $q$-traces,

\begin{equation}\label{nineteen}
c_2= \frac{q}{[2]_q} \{ q^{-2}(tr_qK)^2- tr_q K^2 \} \;,
\end{equation}

\noindent
from which its centrality follows taking into account that, for any integer
$n$, eq. (\ref{ten}) implies that
$K_{2} R_{12} K^n_{1} R^{-1}_{12} = R^{-1}_{21} K^n_{1} R_{21} K_{2}$,
hence $K\, tr_q(K^n)=tr_q(K^n)\, K$. Due to the characteristic equation for
the matrix $K$ \cite{2,16} all other central elements $tr_q(K^n)\;,n>2$
are polynomial functions of $c_1$ and $c_2$.

It should be noted that
due to different factors in the quantum Lorentz group ${\cal L}_{q}$ coaction
(\ref{eight}), the $q$-trace
$c_{1}$, which is identified with the time coordinate $x^{0}$
\cite{9,10,11,12}, is not invariant with respect to
(\ref{eight}) (compare (\ref{eight}) with (\ref{sixteen});
it is, of course, invariant with respect the $SU_{q}(2)$ coaction for which
$M^\dagger =M^{-1}$).
In contrast, the $q$-determinant $c_{2}$ of
$K$ is ${\cal L}_q$-invariant and may
accordingly be used to define the $q$-Minkowski length and metric.
To see its invariance, it is useful
to write it in matrix form \cite{2}, using the $q$-antisymmetrizer (a rank one
projector $P_-(q)$: ${\check R}
= {\cal P} R_{12} = q P_{+} - q^{-1} P_{-} $),

\begin{equation}\label{in-one}
(-1/q)c_2\; P_-=  P_- K_1 \check{R} K_1 P_- \;.
\end{equation}
\noindent
The latter expression transforms under the coaction (\ref{eight})

\begin{equation}\label{in-two}
\begin{array}{ccl}
c'_2\;P_- &=&\phi (c_2)\; P_-= (-q)P_- M_1 K_1 \tilde{M}^{-1}_1
\check{R} M_1 K_1 \tilde{M}^{-1}_1 P_- \\
     &   &     \\& = &(-q) P_- M_1M_2K_1 \check{R}
K_1 \tilde{M}^{-1}_2 \tilde{M}^{-1}_1 P_-
= det_q M det_q \tilde{M}^{-1} c_2 \;P_- \quad,
\end{array}
\end{equation}
\noindent
where a consequence of (\ref{seven}) was used,
\begin{equation}\label{in-three}
\tilde{M}^{-1}_1 \check{R} M_1= M_2 \check{R} \tilde{M}^{-1}_2\,\;,
\end{equation}
\noindent
as well as the matrix definition $(det_q M)\,P_- = P_- M_1 M_2 P_-$ of
the $q$-determinant of the quantum group $GL_q(2)$. Once these
$q$-determinants are set equal to 1 the invariance of $c_2$ follows.

As it was mentioned, using the reality condition
$\tilde{M}^{-1}$=$M^{\dagger}$,
the coaction (\ref{eight}) may be written as $K'=MKM^{\dagger}$. This means
that, since the conjugate (star $\ast$) operation is an
antiautomorphism  ($(ab)^*$=$b^*a^*$, etc.) the reality of the $q$-Minkowski
space may be expressed as in the classical case by the hermiticity of $K$,
$K$=$K^{\dagger}$, a requirement which is consistent
with the coaction (\ref{eight}) and
the RE (\ref{ten}). Indeed, eq. (\ref{ten}) goes to itself after
hermitian conjugation using that $R_{12}^{\dagger}$=$R_{12}^t$=$R_{21}$.
The identification of the REA (\ref{ten}) generators
with the ones of ${\cal M}_{q}$ from \cite{9,10,11} is provided by

\begin{equation}\label{tw}
\left(
\begin{array}{cc}
\alpha & \beta \\
\gamma & \delta \\
\end{array}
\right)
\simeq
\left(
\begin{array}{cc}
qD & B \\
A & C/q \\
\end{array}
\right)
\;.
\end{equation}

\noindent
The comparison with \cite{12} requires introducing the $q$-sigma Pauli
matrices $\sigma^{\mu}_{q}$, which depend on $q$ and are $q$-tensors with
respect to the quantum algebras $su_{q}(2)$ and $sl_{q}(2)$.
This will be discussed in a forthcoming paper \cite{14} in detail.

The algebra of $q$-derivatives ${\cal D}_{q}$,
which is a cornerstone in a definition of a $q$-deformed Poincar\'{e} algebra
\cite{10}, can also be written in a REA form.
To this aim  we could again use the coaction (\ref{eight}) and the reflection
equation (\ref{ten}). Nevertheless, in order to have the simplest
 correspondence with the basis used in \cite{10}, we introduce
a  2$\times$2 matrix  $Y$ satisfying

\begin{equation}\label{twone}
R_{12} Y_{1} R_{12}^{-1} Y_{2} = Y_{2} R^{-1}_{21} Y_{1} R_{21}\;.
\end{equation}

\noindent
The entries of $Y$ are the $q$-derivatives in question. These relations
(\ref{twone})
are invariant with respect to the
following quantum Lorentz group ${\cal L}_q$ coaction:

\begin{equation}\label{twtwo}
Y \longrightarrow Y' = \tilde{M} Y M^{-1}\;.
\end{equation}

\noindent
This invariance is easy to check by multiplying (\ref{twone})
by $\tilde{M}_{2} \tilde{M}_{1}$ from the left  and
by $M_{1}^{-1}M_{2}^{-1}$ from the right and by using (\ref{six}), the
inverse of (\ref{five}) and the transformed (\ref{seven}):

\begin{equation}\label{twthree}
\tilde{M}_{2} R^{-1}_{12} M_{1}^{-1} = M_{1}^{-1} R_{12}^{-1} \tilde{M}_{2}
\;, \qquad
\tilde{M}_{1} R_{21}^{-1} M_{2}^{-1} = M_{2}^{-1} R_{21}^{-1} \tilde{M}_{1}
\quad.
\end{equation}

\noindent
The identification of the algebra ${\cal D}_{q}$ generators with
the derivatives $\partial_{A}
, \partial_{B}, \partial_{C}$, $\partial_{D}$ of \cite{10} is the following

\begin{equation}\label{twfour}
Y \simeq
\left(
\begin{array}{cc}
\partial_{D} & \partial_{A}/q \\
q \partial_{B} & \partial_{C} \\
\end{array}
\right) \;.
\end{equation}
\noindent
The REA (\ref{twone}) gives a compact form for their commutation
properties (see eq. (5.2) in \cite{10}). As in the previous case (cf. eqs.
(\ref{thirteen}), (\ref{fourteen})),
the REA properties \cite{2,5} give the centrality of two elements of
${\cal D}_{q}$

\begin{equation}\label{thtwo}
\partial_{0} \sim \partial_{D} + q^{2} \partial_{C} \quad, \quad
{\Box}_{q} \sim
\partial_{D} \partial_{C} - q^{-2} \partial_{A} \partial_{B}
\end{equation}

\noindent
and the invariance of the latter one, $q$-D'Alembertian operator (cf.
\cite{10}) which may be used to introduce a $q$-Klein-Gordon equation.

The algebra of $q$-derivatives ${\cal D}_{q}$ is isomorphic to the previous
algebra ${\cal M}_{q^{-1}}$ under the map

\begin{equation}\label{twfive}
\sigma_{1} Y \sigma_{1} \simeq K (q^{-1}) \;,
\end{equation}

\noindent
where the properties of the $SL_{q}(2,C)$ matrix $R$ are used

\begin{equation}\label{twsix}
R_{12}^{-1} (q)= R_{12} (q^{-1})=( \sigma_{1} \otimes \sigma_{1}) R_{21}
(q^{-1}) (\sigma_{1} \otimes \sigma_{1}) \;,
\end{equation}

\noindent
$\sigma_{1}$ being the usual Pauli matrix.

The coactions (\ref{eight}) and (\ref{twtwo}) are defined by the same
$q$-Lorentz group. To show this, we have to connect the generators
$M_{ij} (\tilde{M}^{-1})_{kl}$ from (\ref{eight}) with those
$\tilde{M}_{st}(M^{-1})_{mn}$ from (\ref{twtwo}). To this end,
one has to use the defining relations (\ref{seven}) and the
well-known $q$-metric tensor for $GL_q(2)$,

\begin{equation}\label{twseven}
M \varepsilon_q M^t =(det_qM)\; \varepsilon_q \quad,\quad
\varepsilon_q=\left( \begin{array}{cc}
                           0  &  q^{-1/2} \\
                         -q^{1/2}  &  0
                          \end{array} \right)  \quad.
\end{equation}
\noindent
The linear connection among of the different sets of the $q$-Lorentz
generators is

\begin{equation}\label{tweight}
\check{R}^{\varepsilon} M \otimes (\tilde{M}^{-1})^t
(\check{R}^{\varepsilon})^{-1}= \tilde{M} \otimes (M^{-1})^t\;,
\end{equation}
\noindent
where $\check{R}^{\varepsilon}= (\varepsilon^t)_2
\check{R} (\varepsilon^t)^{-1}_2$. Hence both coactions are defined by the
same quantum group.
Moreover, the reality condition for  (\ref{twtwo}),
$M^{-1}$=$\tilde{M}^{\dagger}$, is the same as for (\ref{eight}) since,
although $(M^{-1})^t \neq (M^t)^{-1}$,
$(M^{-1})^{\dagger}$=$(M^{\dagger})^{-1}$.

We have to find now the relations  between
the generators of ${\cal D}_{q}$  and ${\cal M}_{q}$
which are invariant with respect to the
coactions (\ref{eight}), (\ref{twtwo}). Their
commutation relations must be inhomogeneous
since they have to reproduce in the $q$$\rightarrow$1 limit
the classical ones, $\partial_{\mu} x^{\nu}= x^{\nu} \partial_{\mu} +
\delta_{\mu}^{\nu}$. These mixed relations can be written also in the
form of a RE with a constant term (inhomogeneous RE)

\begin{equation}\label{twnine}
Y_{2} R_{12} K_{1} R_{21} =  R_{12} K_{1} R_{12}^{-1} Y_{2} + J \; .
\end{equation}

\noindent
The constant $4 \times 4$ matrix $J$ is fixed (with some ambiguity, see
below) by requiring its invariance

\begin{equation}\label{th}
\tilde{M}_{2} M_{1} J \tilde{M}_{1}^{-1} M_{2}^{-1} = J
\end{equation}

\noindent
and the classical correspondence for $q\rightarrow$1.
{}From (\ref{seven}) one finds

\begin{equation}\label{thone}
J= \eta R_{12} {\cal P} \;,
\end{equation}

\noindent
where the coefficient $\eta$ is set equal to $q^2$ to have the same sixteen
relations as
(5.1) of \cite{10} with  the identifications (\ref{tw}), (\ref{twfour}).
The r.h.s. of (\ref{twnine}) was taken in this way
to simplify the comparison with \cite{10}. But redefining the matrix $Y$
by the factor $q^{\alpha N}$, $Y'$=$q^{\alpha N}Y$, where $N$ is a grading
operator for ${\cal M}_q$ ({\it i.e.}, satisfying $[N,K]$=$K$), it is possible
to obtain any other factor. Multiplying eq. (\ref{twnine})
by $q^{\alpha N}$ it gives

\begin{equation}\label{in-four}
Y'_{2} R_{12} K_{1} R_{21} = q^\alpha R_{12} K_{1} R_{12}^{-1} Y'_{2}
+ q^{\alpha N} J \; ,
\end{equation}

\noindent
where $Y'$ still satisfies (\ref{twone}).
Although the inhomogeneous term is modified, eq. (\ref{in-four}) still
provides the desired classical limit. We note that the consistency of
the defining relations (\ref{seventeen}, \ref{twone}, \ref{twnine}) with
associativity, which here could be checked using only the previous
matrix formalism and the YBE for the $R$ matrix (\ref{three}),
follows from the correspondence with \cite{10}.

As we have seen, we {\it may} have the same reality conditions
for ${\cal M}_q$  and ${\cal D}_q$ ({\it i.e.}, $K^{\dagger}$=$K$,
$Y^{\dagger}$=$Y$). It should be noticed,
however,  that more elaborated, non-linear expressions still respecting
the covariance (as for instance $Y^{\dagger}$=$Y + \lambda YKY$)
are also possible, although their consistency with the defining
relations (\ref{twone}, \ref{twnine}) is not {\it a priori} evident.
As for the hermitian conjugation, the
inhomogeneous equation (\ref{twnine}) is not invariant.
As a result, the $\ast$ operation   for ${\cal D}_q$
is up to now a rather elaborated one and the conjugated $q$-derivatives
are expressed non-linearly in terms of $Y$
and $K$ according to \cite{10}.

It is also possible to introduce in the
framework of non-commutative geometry on $q$-Minkowski space
an algebra of $q$-one-forms with generators $d\alpha,...,d\delta$
($dK$  in matrix form),
where $d$  is the exterior derivative with the usual
properties  of nilpotency, linearity and non-deformed Leibniz rule \cite{10}.
The corresponding commutation relations
among $q$-one-forms and `coordinates'
(the sixteen relations (B.1) and ten (B.5), (B.6) of \cite{10}), as well
as those with the $q$-derivatives, can be written as RE for $K$ and $dK$,
$dK$ and $dK$, $Y$ and $dK$ as
\begin{equation}\label{in-five}
\begin{array}{rcl}
R_{12} K_{1} R_{21} dK_{2} &=& dK_{2} R_{12} K_{1} R_{12}^{-1}\;, \\
        &   &    \\
R_{12} dK_{1} R_{21} dK_{2} &=& -dK_{2} R_{12} dK_{1} R_{12}^{-1}\;, \\
      &   &    \\
Y_2 R_{21}^{-1} dK_1 R_{21} &=& R_{12} dK_1 R_{21} Y_2  \quad .
\end{array}
\end{equation}
\noindent
while the expression for the exterior derivative operator can be given
in terms of the $q$-trace (\ref{fifteen}),
\begin{equation}\label{in-six}
d=q^{-1}\,tr_q\,(dK)\,Y\,.
\end{equation}
\noindent
It is worth mentioning that it is possible to replace in  each side
of equations
(\ref{ten}), (\ref{twone}), (\ref{twnine}) and (\ref{in-five}) {\it one}
of the matrices $R_{ij}$ by $R_{ji}^{-1}$. By doing so, a second
consistent covariant calculus may be introduced which reproduces the
formulae (B.3), (C.5) and (C.7) of \cite{10} related to the conjugated
generators, but we shall not
discuss this further here.

To conclude, we would like to make some general comments and remarks.
{}From a physical point of view, it is important to discuss the
applications of the $q$-deformation of a `classical' theory.
Finding {\it the} deformation is already a difficulty in itself, since in many
instances there is not a unique prescription to $q$-deform a classical
structure. This is the case for the  inhomogeneous Lie groups, for which
the Drinfel'd-Jimbo or FRT prescriptions,
valid for simple groups, are absent. Also,
for quantum groups there are more possibilities for $q$-homogeneous
spaces than there are for Lie groups (this is the case for the $SU_q(2)$
$q$-spheres). In the present discussion of $q$-Minkowski space
and related algebras, we have also encountered a few relations which allow
for different, consistent expressions with the same $q\rightarrow$1 limit.

 From the
point of view of exploiting the covariance properties of the algebras
treated in this paper, it would be more appropriate to discuss the quantum
algebra aspects rather than the quantum group ones, since in the first
instance there is a finite set of independent non-commuting generating
elements (the matrix elements of $K$, $Y$, etc) while the entities of the
second are new ones unrelated to them. For instance, the $q$-relativistic
invariant operators constitute an example of a direct
application of the ${\cal M}_q$
and ${\cal D}_q$ algebras once a proper basis has been chosen
(see, {\it e.g.}
the invariant $q$-Klein-Gordon operator in (\ref{thtwo}); a
$q$-Lorentz covariant $q$-Dirac operator is obtained similarly).
The definition of the $q$-Lorentz algebra $l_q$
generators and their action on the
elements of ${\cal M}_{q}$ or, better, their commutation relations with
generators of ${\cal M}_{q}$ and ${\cal D}_{q}$,
could be given by the duality of the Hopf algebras ${\cal L}_q\sim (l_q)^*$
and the coactions (\ref{eight}), (\ref{twtwo}).

Coming back to the physical problems, the simplest application of
a $q$-Minkowski space would be the `$q$-relativistic' particle.
The multiparticle interpretation requires endowing the ${\cal M}_{q}$ algebra
with the structure of a Hopf algebra or a twisted (braided) Hopf algebra (cf.
\cite{10,5,15,16}). Finally, we wish to mention that the essential ingredients
for implementing a coaction consistent with an associative non-commuting
 algebra, eqs. (\ref{eight}, \ref{ten}) and their analogous, are also valid
for  other $q$-groups defined through (\ref{five}, \ref{six}, \ref{seven}).
This and some of the previously mentioned extensions will be presented
elsewhere \cite{14}.

\vspace{4\baselineskip}

\noindent
{\bf Acknowledgements:} This paper is partially supported by a CICYT
research grant.
We wish to thank D. Ellinas and R. Sasaki for helpful
discussions and comments. Two of us (P.K. and F.R.) also wish to thank
the Spanish Ministry of Education and Science for financial support.

\end{document}